\documentclass[a4paper, 12pt, openany, oneside]{article}
\usepackage[cp1251]{inputenc} %- для WiN
\usepackage[english,russian]{babel} %- для рус-англ. переноса
\usepackage{ graphics,amsfonts, amsmath,bm,amssymb, amsmath}
\usepackage[dvips]{graphicx}
\usepackage[flushleft,small]{caption2} % чтобы в подписях к рисункам
\renewcommand{\@biblabel}[1]{#1.\hfill}
\newcommand{\diag}{\rm \diag\, }

\renewcommand{\Re}{\mathop{\rm Re\,}}

\newcommand{\Res}{\mathop{\rm Res\,}}
\makeatother {

\begin{document}
 \thispagestyle{empty}
\large
\renewcommand{\refname}{\normalsize\begin{center}\rm
LITERATURE\end{center}}

\begin{center}
SKIN EFFECT PROBLEM WITH THE DISPLACEMENT CURRENT IN MAXWELL PLASMA
BY THE SOURCE METHOD
\end{center}

\begin{center}
 {\bf Y.F. Alabina, A.V. Latyshev, A.A. Yushkanov}
\end{center}

\begin{center}
 Moscow State Regional University\\
{\it 105005, Moscow, Radio st., 10 а
\\e-mail: yf.alabina@gmail.com, avlatyshev@mail.ru,
yushkanov@inbox.ru}
\end{center}

%\begin{abstract}
We found the analytical solution to the problem of the skin effect
for Maxwell plasma with the use of the kinetic equation, where the
frequency of electron collisions is constant. We use the specular
reflection of electrons from the surface as a boundary condition.
The behavior of an impedance near to a plasma resonance is
considered. We consider limiting cases of skin effect.

{\bf Introduction.} Skin effect is the plasma response to variable
electromagnetic field, tangential to the surface \cite{1a, 1}.

First, the analytical solution of the skin effect problem at any
value of anomaly parameter was found in \cite{2} and \cite{3} for
plasma in metal. For the gas plasma, the corresponding solution is
considered in \cite{4}. There has been substantial interest to this
problem \cite{5,6,7,8,9}. In \cite{4}, the behavior of plasma near
the resonance is not considered. Also, research of impedance with
displacement current near to resonance isn't carried out in the
previous works. For example, the displacement current isn't taken
into account  in [2]. Research of surface impedance near to plasma
resonance isn't carried out in the rest of works.  We research the
behaviour of impedance near to plasma resonance with the
displacement current in this paper.

In this paper we continue the development of an analytical method of
solving boundary problem for systems of  equations for the electric
field in half-space gas plasma. The basis of the method is the idea
of the symmetrical continuation of the electric field to the
conjugate half-space. We provide the analytical solution of the
boundary problem of the skin effect theory for electron plasma, that
fills the half-space. We formulate analytical expressions for
electric field, distribution function of electrons and impedance.

We assume that electromagnetic wave is incident normally to the
interface of the plasma. In such configuration, the electric field
of electromagnetic wave has only tangential component. We use the
specular electron reflection from interface as boundary condition.
The interface of ions on the conductivity of plasma is not
considered.

 {\bf 1. Problem statement and basic equations.}

Let Maxwell plasma fills the half-space $x>0$, where $x$ is the
coordinate orthogonal to plasma boundary. Let the external electric
field has only  $y$ component. Then the self-concordance electric
field inside in plasma also has only $y$ component
$E_y(x,t)=E(x)e^{-i\omega t}$. Let us take the kinetic equation for
distribution function of electrons:

$$
\dfrac{\partial f}{\partial t}+\text{v} _{x}\dfrac{\partial
f}{\partial x}+eE(x)e^{-i\omega t}\dfrac{\partial f}{\partial
p_y}=\nu(f_0-f(t,x,\mathbf{v})). \eqno{(1)}
$$

In (1) $\nu$ is the frequency of electron collisions with ions,
$e_0$ is the charge of electron, $f_0(\text{v})$ is the equilibrium
Maxwell distribution function,
$$
f_0(\text{v})=n\left(\dfrac{\beta}{\pi}\right)^{3/2}\exp(-\beta^2\text{v}^2),\quad
\beta=\dfrac{m}{2k_BT}.
$$

Here  $m$ is the mass of electron, $k_B$ is the Boltzmann constant,
$T$ is the temperature of plasma, $\text{v}$ is the velocity of the
electron, $n$ is the concentration of electrons, $c$ is the speed of
light.

The electric field $E(x)$ satisfies Poisson's equation
$$
E''(x)+\dfrac{\omega^2}{c^2}E(x)= -\dfrac{4\pi i  e^{i\omega
t}\omega e}{c^2} \int v_y f(t,x,\mathbf{v})\,d^3v. \eqno{(2)}
$$

Assume that the intensity of an electric field is such that linear
approximation is valid. Then distribution function can be
represented in the form:
$$
f=f_0\left(1+C_y\exp(-i\omega t)h(x,\mu)\right),
$$
where $\textbf{C}=\sqrt{\beta}\text{v}$ is the dimensionless
velocity of electron, $\mu=C_x$. Let $l=v_T\tau$ be the mean free
path of electrons, $v_T=1/\sqrt{\beta},\; \tau=1/\nu$. We introduce
the dimensionless values:
$$
t_1=\nu t,\quad x_1=\dfrac{x}{l}, \quad
e(x_1)=\dfrac{\sqrt{2}e}{\nu\sqrt{mk_BT}}E(x_1).
$$

Below, instead of $x_1$ we shall write again $x$. In new variables,
the kinetic equation (1) and the equation on a field with the
 displacement current (2) become
$$
\mu\dfrac{\partial h}{\partial x}+z_0\,h(x,\mu)=e(x),  \quad
z_0=1-i\omega\tau, \eqno{(3)}
$$
$$
%\dfrac{d^2e(x)}{d\,x^2}
e''(x)+Q^2e(x)=-i\dfrac{\alpha}{\sqrt{\pi}}
\int\limits_{-\infty}^{\infty}\exp(-{\mu'}^2)\,h(x,\mu')\,d\mu',\quad
Q=\dfrac{\omega l}{c}, \eqno{(4)}
$$
where $\delta=\dfrac{c^2}{2\pi\omega\sigma_0}$ is the classical
depth of the skin layer, $ \quad \sigma_0=\dfrac{e^2n}{m\nu},$
$\alpha=\dfrac{2l^2}{\delta^2}$, what $\alpha$ as the anomaly
parameter.

We formulate the boundary conditions for the distribution function
of the electron in case of the specular electron reflection from the
surface:
$$
h(0,\mu)=h(0,-\mu), \qquad 0<\mu<+\infty. \eqno{(5)}
$$
We use the condition that function $h(x,\mu)$ vanishes far from the
surface:
$$
h(+\infty,\mu)=0, \qquad -\infty<\mu<+\infty, \eqno{(6)}
$$
and conditions for electric field on the interface and far from it:
$$
{e}'(0)={e_s}', \qquad e(+\infty)=0, \eqno{(7)}
$$
where ${e_s}'$ is the given value of gradient of electric field on
the plasma interface.

So, the skin effect problem is formulated completely. We seek
solution of system of the equations (3) and (4) in this problem that
satisfy boundary conditions (5)--(7).

{\bf 2. The analytical solution of the problem.} As a first step in
the source method, we extend the electric field and distribution
function to the "negative"\, half-space $x<0$:
$$
e(x)=e(-x), \qquad h(x,\mu)=h(-x,-\mu). \eqno{(8)}
$$
After we substitute $x=0$ to equation (8), we obtain that electric
field and distribution function of electrons are continuous and the
derivative of an electric field has discontinuity:
$e'(+0)-e'(-0)=2{e_s}'$. In account of this circumstance, we
introduce term with Dirac delta function to the field equation
\cite{9}:
$$
%\dfrac{d^2e(x)}{d\,x^2}
e''(x)+Q^2e(x)-2e_s'\delta(x)=%$$$$=
-i\dfrac{\alpha}{\sqrt{\pi}}
\int\limits_{-\infty}^{\infty}\exp(-{\mu'}^2)\,h(x,\mu')\,d\mu',
\eqno{(9)}
$$
where $\delta(x)$ is the Dirac delta function.

The third term in the left hand side of the equation (9) corresponds
to discontinuity of a derivative of the electric field for $x=0$.

 The solution of the (3), (9), (5)--(7) can be sought as Fourier integrals (by variable $x$):
$$
e(x)=\dfrac{1}{2\pi}\int\limits_{-\infty}^{\infty}e^{ikx}E(k)\,dk,
\eqno{(10)}
$$
$$
h(x,\mu)=\dfrac{1}{2\pi} \int\limits_{-\infty}^{\infty}e^{ikx}
\Phi(k,\mu)\,dk, \eqno{(11)}
$$
$$
\delta(x)=\dfrac{1}{2\pi}\int\limits_{-\infty}^{\infty} e^{ikx}\,dk.
\eqno{(12)}
$$
We substitute (10)--(12) into (3) and (9). We get the following
 system of the characteristic equations:
$$
\Big(Q^2-k^2\Big)E(k)-2e'(0)=-i\dfrac{\alpha}{\sqrt{\pi}}
\int\limits_{-\infty}^{\infty}\exp(-\mu^2)\Phi(k,\mu)\,d\mu,
$$
$$
\Phi(k,\mu)(z_0+ik\mu)=E(k).
$$
From these equations, we get spectral densities of the distribution
function and electric field, respectively:
$$
\Phi(k,\mu)=\dfrac{E(k)}{ik\mu+z_0}, \eqno{(13)}
$$
$$
E(k)=-\dfrac{2 e_s'}{k^2\lambda(k)}, \eqno{(14)}
$$
where
$$
\lambda(k)=1-\dfrac{Q^2}{k^2}-i\dfrac{\alpha}{k^2\sqrt{\pi}}
\int\limits_{-\infty}^{\infty}\dfrac{\exp(-\mu^2)\,d\mu}{ik\mu+z_0}.
%\eqno{(4.14)}
$$

To find the profile of the electric field in the half-space, we
substitute (14) into (10):
$$
e(x)=-\dfrac{e_s'}{\pi}\int\limits_{-\infty}^{\infty}
\dfrac{e^{ikx}\,dk}{k^2\lambda(k)}.\eqno{(15)}
$$

To get the function of the electron distribution in the half-space
we substitute (14) into (13). It is obvious that this spectral
density is:
$$
\Phi(k,\mu)=-\dfrac{2e_s'}{(z_0+ik\mu)k^{2}\lambda(k)}. \eqno{(16)}
$$

Now we substitute (16) into (11). We get
$$
h(x,\mu)=-\dfrac{e_s'}{\pi}
\int_{-\infty}^{\infty}\dfrac{e^{ikx}dk}{(z_0+ik\mu) k^2\lambda(k)}.
%\eqno{(2.10)}
$$

{\bf 3. The impedance evaluation.} We introduce the dimensionless
decrease of the electric field into the depth of plasma:
$$
\Lambda(\alpha,\Omega)=-\dfrac{e(0)}{e_s'}, \qquad
\Omega=\omega\tau. \eqno{(17)}
$$

From (15) this dimensionless decrement is:
$$
\Lambda(\alpha,\Omega)=\dfrac{1}{\pi}\int\limits_{-\infty}^{\infty}
\dfrac{dk}{k^2\lambda(k)}=\dfrac{2}{\pi}\int\limits_{0}^{\infty}
\dfrac{dk}{k^2\lambda(k)}.
%\eqno{(3.1)}
$$

The value of the impedance can be calculated with the help of the
formula (see \cite{1a}):
$$
Z=\dfrac{4\pi i \omega}{c^2}
\dfrac{e(x)}{\left.\dfrac{de(x)}{dx}\right|_{x=0}}.
$$

It should be noted that
$$
\dfrac{de(x')}{dx'}=\dfrac{de(x)}{dx}\cdot \dfrac{dx}{dx'}=
%\dfrac{1}{\nu \sqrt{\beta}}\cdot \dfrac{de(x)}{dx}=
l\dfrac{de(x)}{dx}.
$$

Thus, according to the previous formula the impedance is
$$
Z=\dfrac{4\pi i \omega l}{c^2}\cdot \dfrac{e(0)}{e_s'},
$$
or, taking into consideration the equality (17),
$$
Z=-i\dfrac{4\pi \omega l}{c^2}\Lambda(\alpha,\Omega). \eqno{(18)}
$$

Let us introduce parameter $R$, which is equal to the modulus of
impedance in normal skin effect (when $\alpha \ll 1,\; \Omega \ll
1$)
$$
R=\sqrt{\dfrac{4\pi \omega}{c^2 \sigma_0}},
$$
where $\sigma_0$ is the static electrical conductivity of plasma
(for $\omega=0$). Now formula (18) for the impedance can be written
in the form $Z=RZ_0$, where $Z_0$ is the dimensionless part of
impedance,
$$
Z_0=-i \sqrt{\alpha}\Lambda(\alpha,\Omega). \eqno{(19)}
$$

{\bf 4. The analysis of the solution.} We substitute the variable
$k=1/t$ to the integral (19). In this case the dimensionless
decrement is:
$$
\Lambda(\alpha,\Omega)=\dfrac{2}{\pi}\int\limits_{0}^{\infty}
\dfrac{dt}{\lambda(1/t)}.
$$

Here
$$
\lambda(1/t)=1-Q^2t^2-\dfrac{\alpha t^3}{\sqrt{\pi}}
\int\limits_{-\infty}^{\infty}\dfrac{\exp(-\mu^2)\,d\mu}{\mu-
iz_0\,t}.
%\eqno{(4.21)}
$$

We will present special partial cases of formula (19).

We will start from the case of the normal skin  effect. In this case
$$
\alpha\ll 1,\;\qquad \Omega\ll 1,\;\qquad z_0=1-i\omega\approx 1.
$$
We suppose that there is no current of the bias. In this case the
dimensionless impedance depends on $\alpha$ is
$$
Z_0=-i\sqrt{\alpha}\,\dfrac{2}{\pi}\int\limits_{0}^{\infty}
\dfrac{dt}{1-\alpha t^3\dfrac{1}{\sqrt{\pi}}
\int\limits_{-\infty}^{\infty}\dfrac{\exp(-u^2)\,du}{u-it}},
$$
or, if we introduce
$$
t_0(iz)=\dfrac{1}{\sqrt{\pi}}\int\limits_{-\infty}^{\infty}
\dfrac{\exp(-u^2)\,du}{u-iz},
$$
the expression for the dimensionless impedance will be  follows:
$$
Z_0=-i\sqrt{\alpha}\;\dfrac{2}{\pi}\int\limits_{0}^{\infty}
\dfrac{dt}{1-\alpha t^3\,t_0(it)}.
%\eqno{(4.22)}
$$

It should be noted that for the large $t$: $t_0(it)\approx i/t$.
Therefore
$$
Z_0=-i\sqrt{\alpha}\;\dfrac{1}{\pi}\int\limits_{-\infty}^{\infty}
\dfrac{dt}{1-i\alpha t^2}= 2\sqrt{\alpha}\;\cdot\Res_{z=-\frac{1-i}
{\sqrt{2\alpha}}} \dfrac{1}{1-i\alpha \,z^2}=\dfrac{1-i}{\sqrt{2}}.
$$
The expression $Z_0=\dfrac{1-i}{\sqrt{2}}$ is a well-known classical
result \cite{4}.

We will consider the anomalous skin effect in the low-frequency
limit, that is, when $\alpha \gg 1,\; \Omega\ll 1, \;z_0\approx 1$.
In this case for small $t$ we have:
$$
t_0(it)\approx \dfrac{1}{\sqrt{\pi}}\int\limits_{-\infty}^{\infty}
\dfrac{\exp(-u^2)\,du}{u-it}\approx i\sqrt{\pi}\exp(-(it)^2)\approx
i\sqrt{\pi}.
$$
Thus the non dimensional impedance is
$$
Z_0=-i\sqrt{\alpha}\dfrac{2}{\pi}\int\limits_{0}^{\infty}
 \dfrac{dt}{1-i\alpha \sqrt{\pi}t^3}.
$$
We substitute the variable in this integral $t=1/k$ and obtain
$$
Z_0=-i\sqrt{\alpha}\dfrac{2}{\pi}\int\limits_{0}^{\infty}
\dfrac{k\,dk}{k^3-i\alpha \sqrt{\pi}}.
$$
We substitute one more variable $k=\sqrt[3]{\alpha \sqrt{\pi}}x$, so
$$
Z_0=-i\sqrt{\alpha}\dfrac{2\sqrt[3]{\alpha \sqrt{\pi}}}{\pi}
\int\limits_{0}^{\infty}\dfrac{x\;dx}{x^3-i}.
$$
We consider the integral
$$
J=\int\limits_{0}^{\infty}\dfrac{x\;dx}{x^3-i}=J_1+iJ_2,\quad
J_1=\int\limits_{0}^{\infty}\dfrac{x^4\;dx}{x^6+1},\quad
J_2=\int\limits_{0}^{\infty}\dfrac{x\;dx}{x^6+1}.
$$

We will calculate the integral $J_1$ with the use of residues. The
function under integral has simple poles in the points
$z_k=\exp\Big(\dfrac{i\pi}{6}(1+2k)\Big),\; k=0,1,2$. So
$$
J_1=\dfrac{1}{2}\int\limits_{-\infty}^{\infty}\dfrac{x^4dx}{x^6+1}=
\pi i\;\sum_{k=0}^1\Res_{z=z_k}\;\dfrac{z^4}{z^6+1}= \dfrac{\pi
i}{6} \Big(\dfrac{1}{z_0}+\dfrac{1}{z_1}+\dfrac{1}{z_2}\Big)=
\dfrac{\pi}{3}.
$$
The second integral is calculated by the decomposition of the
function under integral into elementary fractions. As a result we
have: $J_2=\dfrac{\pi}{3\sqrt{3}}$. Thus,
$J=\dfrac{\pi(\sqrt{3}+i)}{3\sqrt{3}}$, and the expression for the
non dimensional impedance is
$$
Z_0=\dfrac{2\;\sqrt[6]{\alpha}}{3\sqrt{3}}(1-i\sqrt{3}),
$$
which also coincides with the classical result \cite{4}.

Further, to study the impedance near to a plasma resonance, it is
convenient to use the following dimensionless parameters:
$$
\gamma=\dfrac{\omega}{\omega_p}, \qquad \varepsilon=\dfrac{\nu}
{\omega_p},\qquad \text{where}\quad\omega_p=\dfrac{4\pi e_0^2 n}{m}.
$$
Here $\omega_p$ is the plasma frequency.

We express the parameters of the problem $\alpha, \Omega, Q$ through
$\gamma$, $\varepsilon$ and $v_c=v_T/c$, where $v_T=1/\sqrt{\beta}$
is the heat velocity of the electrons. We obtain that
$$
\alpha=\dfrac{\gamma v_c^2}{\varepsilon^3}, \quad Q=\dfrac{\gamma
v_c}{\varepsilon}, \quad \Omega=\dfrac{\gamma}{\varepsilon}.
$$

We analyze numerically the growth of value of the modulus of
impedance, the real, imaginary parts of impedance and argument of
impedance depending on change of value $\gamma$ from $0.5$ to $1.2$
with various values of other parameters. If we change of the anomaly
parameter $\alpha$ in the specified limits, the value $\gamma$
becomes $\gamma=1$, that is $\omega=\omega_p$ i.e. the oscillation
frequency of external field is the value of plasma frequency. This
is plasma resonance. It would be interesting to consider parameters
of self-concordant field near plasma resonance. In conclusion we
show the results of numerical analysis.

%\newpage
{\bf Conclusion.} The analysis of plots in figure 1a shows that at
the same temperature of plasma, the maximum of the modulus of
impedance is reached at $\gamma=1$, i.e. for $\omega=\omega_p$.
Thus, the less is the effective frequency of electrons collisions
with particles of plasma, the greater is the modulus of impedance.

From figure 1b we see that at the same frequency of collisions of
electrons the maximum of the modulus of impedance is reached at
$\omega=\omega_p$, independent of the temperature. Actual curves in
this figure are computed for various values of parameter
$v_c=v_T/c$, which depends on temperature (it is proportional to
root square of temperature): $v_c=\sqrt{2k_BT/c^2m}$. Thus  the size
of the modulus of impedance decreases quickly with growth of
temperature.

It is interesting to note that for decrease reduction of value
$\varepsilon$ from $10^{-2}$ to $10^{-4}$ (by two orders), the
modulus of impedance also increases  by two orders, more precisely
$95$ times. The temperature of the plasma here is $3000 K$. If
temperature of the plasma is $5000 K$, the modulus of impedance
increases by $97$ times for the same reduction in the value of
$\varepsilon$.

For an increase in the  temperature of the plasma from $3000 K$
($v_c=10^{-3}$) to $5000 K$ ($v_c=13\cdot 10^{-3}$) the value of the
modulus of impedance increases by $170$ times, and for a change in
the temperature of the plasma from $1000 K$ to $3000 K$, the value
of the modulus of impedance changes only by a factor of $2.8$. Thus,
the growth of the modulus of impedance depends on the nonlinear
change in temperature.

In figures 2a and 2b we show the plots of the real part of impedance
(more precisely, the plots of values $\Re(-Z_0)$). The analysis of
plots in figures 2 shows that as the frequency of collisions of
electrons increases, the value $\Re(-Z_0)$ grows at a constant
temperature.If the frequency of collisions of electrons is
constant,then this value grows as the temperature increase.

Let us note, that in figures 1 and 2 we used the logarithmic scale
for vertical axes.

The analysis of dependence of argument of impedance on parameter
$\gamma$ shows that near to plasma resonance, the argument of
impedance has step irrespective of the frequency of electron
collisions (fig. 3) and from temperature.

So, the numerical analysis of plots (fig. 1, 2) shows that near to
plasma resonance, the modulus and imaginary part of impedance have
the sharp maximum, which is absent in low-frequency limit, or in the
normal skin effect theory, where the argument near to resonance has
step, and the real part of impedance has the sharp maximum.

\begin{figure}[h]
\begin{center}
\includegraphics[width=0.45\textwidth]{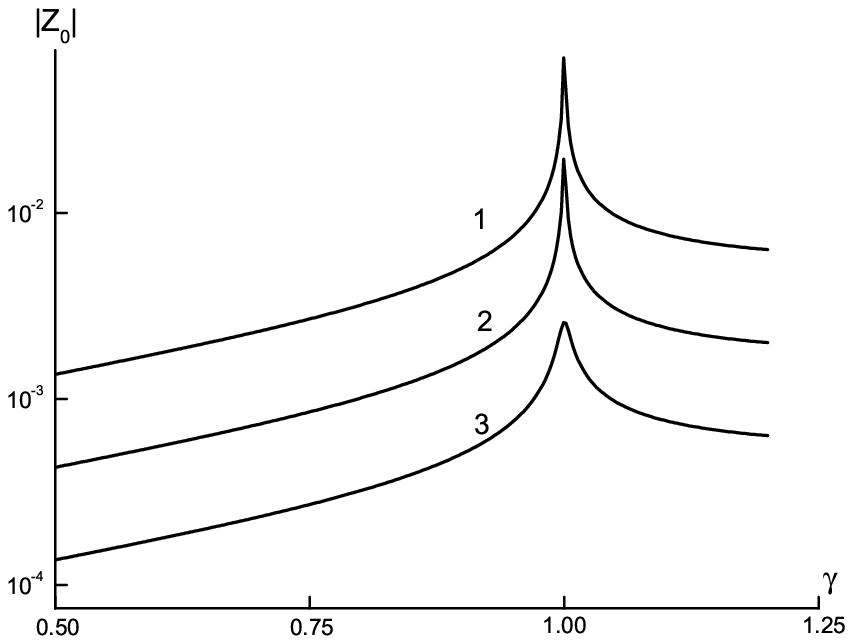}
\hfill
\includegraphics[width=0.45\textwidth]{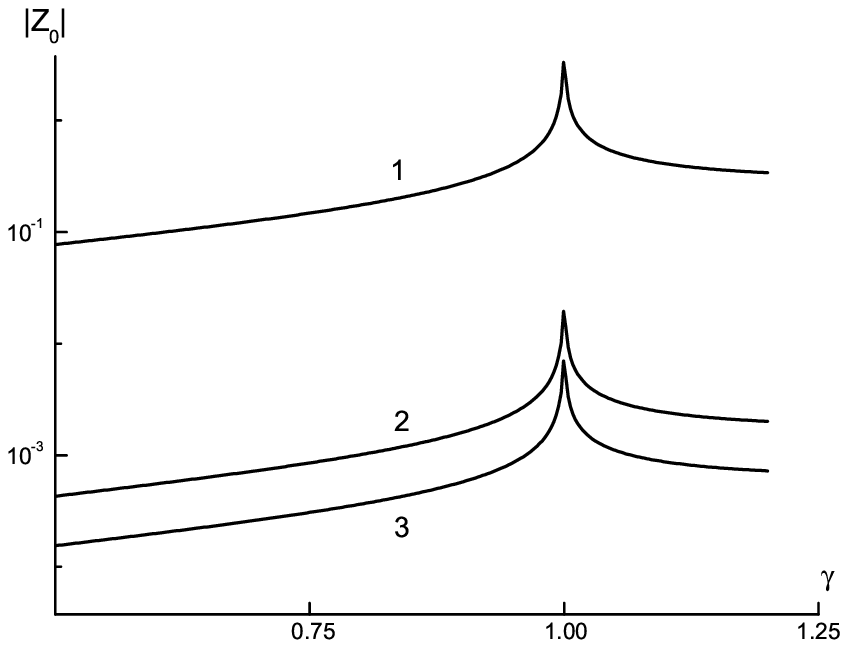}
\\
\parbox[t]{0.47\textwidth}{Fig. 1a. The modulus of impedance. For the curves
$1,2,3$, $\varepsilon=10^{-4}, 10^{-3}, 10^{-2}$ respectively;  and
$v_c=10^{-3}$.} \hfill
\parbox[t]{0.47\textwidth}{Fig. 1b. The modulus of impedance. For the curves
$1,2,3$, $v_c=6\cdot 10^{-4}$, $10^{-3}$, $13\cdot 10^{-3}$
respectively; and $\varepsilon=10^{-3}$.}
\end{center}
\end{figure}

\begin{figure}[h]
\begin{center}
\includegraphics[width=0.45\textwidth]{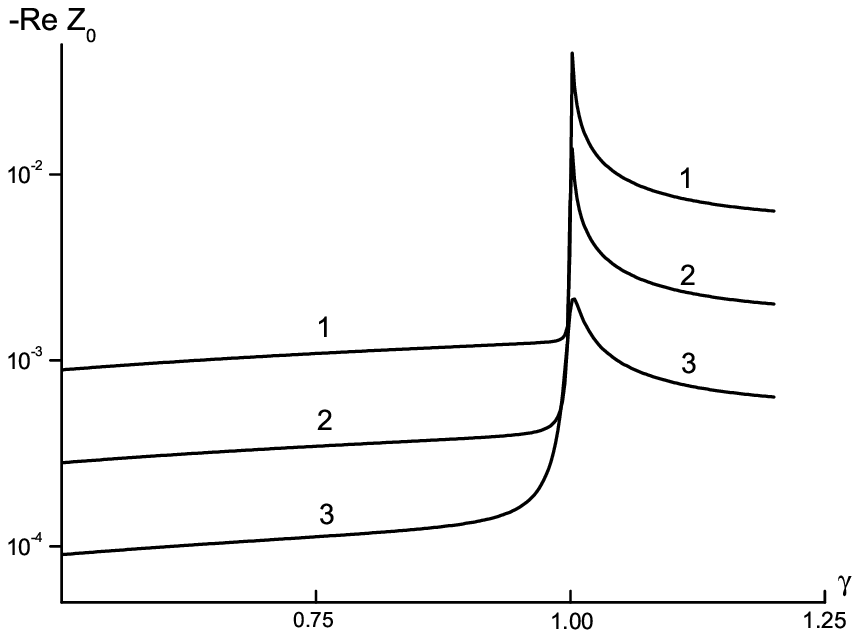}
\hfill
\includegraphics[width=0.45\textwidth]{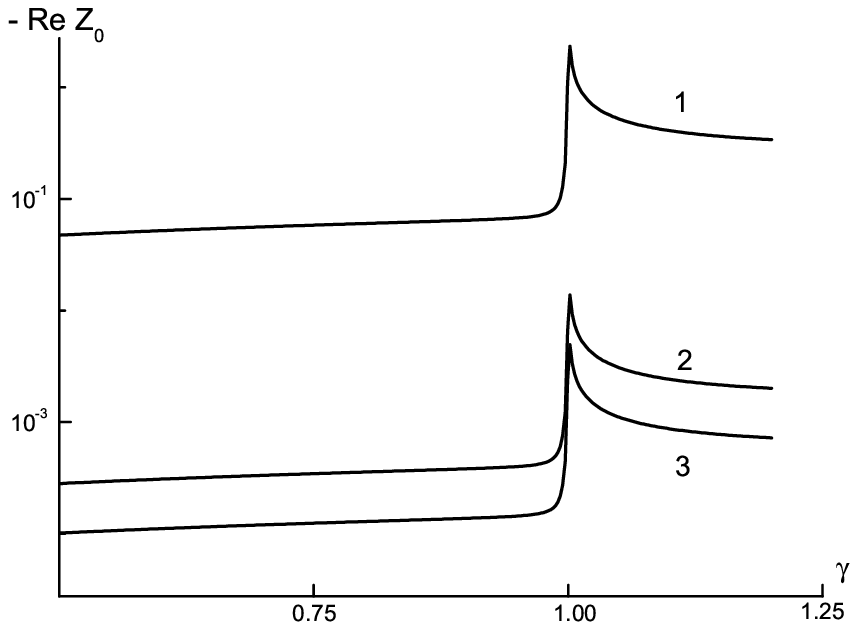}
\\
\parbox[t]{0.47\textwidth}{Fig. 2a. The real part of impedance.
For the curves $1,2,3$, $\varepsilon=10^{-4}, 10^{-3}, 10^{-2}$
respectively; and $v_c=10^{-3}$.} \hfill
\parbox[t]{0.47\textwidth}{Fig. 2b. The real part of impedance.
For the curves $1,2,3$, $v_c=13\cdot 10^{-3}$, $10^{-3}$, $6\cdot
10^{-4}$ respectively; and $\varepsilon=10^{-3}$.}
\end{center}
\end{figure}

\begin{figure}[h]
\begin{center}
\includegraphics[height=0.2\textheight]{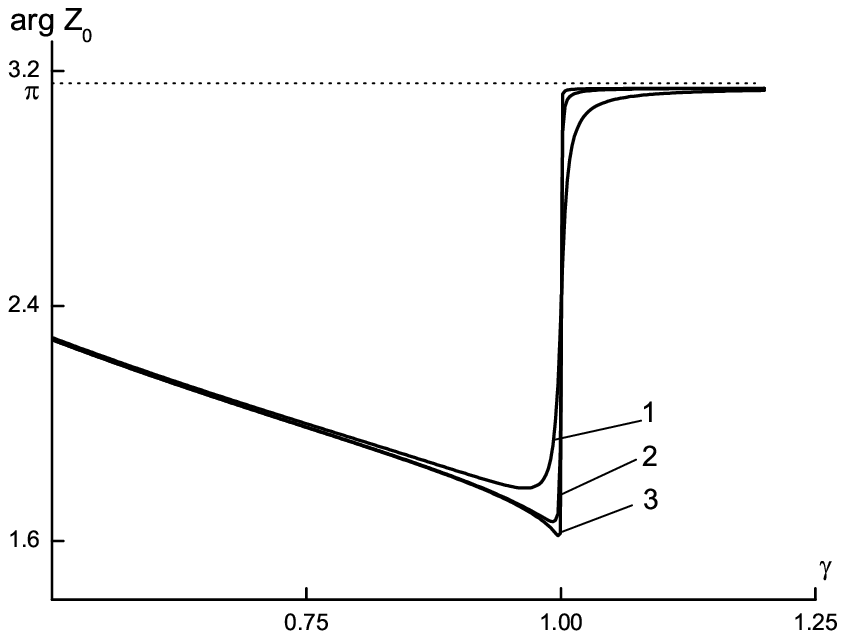}
\end{center}
\begin{center}
  Fig. 3. The argument of impedance. For the curves $1,2,3$,
$\varepsilon=10^{-2}$, $10^{-3}$, $10^{-4}$ respectively;  and
$v_c=10^{-3}$.
\end{center}
\end{figure}

\end{document}